\documentclass[twocolumn,prl,aps,showpacs,groupedaddress,byrevtex]{revtex4}
\usepackage{graphics}
\usepackage{amsmath}

\newcommand{\im}{\mathrm{ i}}
\newcommand{\x}{\vec{x}}
\newcommand{\q}{\vec{q}}
\newcommand{\xp}{\vec{x}'}

\begin{document}
\title{Experimental evidence of high-resolution ghost imaging and ghost diffraction with classical thermal light}

\author{D. Magatti} 
\affiliation{INFM, Dipartimento di Fisica e Matematica, Universit{\`a} dell'Insubria,
  Via Valleggio 11, 22100 Como, Italy}
\author{F. Ferri} 
\affiliation{INFM, Dipartimento di Fisica e Matematica, Universit{\`a} dell'Insubria,
  Via Valleggio 11, 22100 Como, Italy}
\author{A. Gatti}
\affiliation{INFM, Dipartimento di Fisica e Matematica, Universit{\`a} dell'Insubria,
  Via Valleggio 11, 22100 Como, Italy}
\author{M. Bache} 
\affiliation{INFM, Dipartimento di Fisica e Matematica, Universit{\`a} dell'Insubria,
  Via Valleggio 11, 22100 Como, Italy}
\author{E. Brambilla}
\affiliation{INFM, Dipartimento di Fisica e Matematica, Universit{\`a} dell'Insubria,
  Via Valleggio 11, 22100 Como, Italy}
\author{L.A. Lugiato} 
\affiliation{INFM, Dipartimento di Fisica e Matematica, Universit{\`a} dell'Insubria,
  Via Valleggio 11, 22100 Como, Italy}

 \date{\today}
\begin{abstract}
  High-resolution  ghost image and ghost
  diffraction  experiments are performed by using a single source of 
thermal-like speckle light divided by  a
  beam splitter. Passing from the image to the diffraction result
  solely relies on changing the optical setup in the reference arm,
  while leaving untouched the object arm. The product of spatial resolutions
of the ghost image and ghost diffraction experiments is shown to overcome a limit which was formerly thought to be achievable only with entangled photons.
\end{abstract}

\maketitle

The ghost imaging protocol, 
provides a flexible way of
performing coherent imaging by using spatially incoherent light.
It relies on the use of two spatially correlated beams, one of them illuminating an object, while the other holds a known reference optical setup. Information on the object spatial distribution is obtained
by correlating the spatial distributions of the
two beams. 
Traditionally parametric down-conversion (PDC) is the source of the
correlated beams. In
the low-gain regime single pairs of  entangled signal-idler photon  can be resolved, and the
information is extracted from coincidence measurements
\cite{pittman:1995,strekalov:1995,ribiero:1994,abouraddy:2001,abouraddy:2002},
while in the high-gain regime several photon pairs form entangled
beams and the information is contained in the
signal-idler intensity correlation \cite{gatti:2003,gatti:2004,bache:2004,bache:2004a}.
Landmark experiments in the low-gain regime showed that using 
entangled photons both the object image could be retrived (the ``ghost
image'' experiment \cite{pittman:1995}) as well as the object
diffraction pattern (the ``ghost diffraction'' experiments
\cite{strekalov:1995,ribiero:1994}). 
\par
Recently, a very lively debate arised, aiming to identify which aspects
(if any) of ghost imaging truly requires entaglement and which could be reproduced by using classically correlated sources.
The first theoretical interpretation of the experiments suggested that
entanglement of photon pairs was essential to retrieve information from the
correlations\cite{abouraddy:2001,abouraddy:2002}.
This interpretation was challenged both by theoretical arguments \cite{gatti:2003} and experiments \cite{bennink:2002a,bennink:2004},
which showed that virtually any single result of ghost imaging could be reproduced by using classical sources with the proper kind of spatial correlation.
However, theoretical investigations showed that
only quantum entanglement provides perfect correlations in both
photon position (near-field) and momentum (far-field)   \cite{gatti:2003,brambilla:2004}, and suggested that this 
is crucial if both the image and the diffraction pattern are to be
retrieved from the same source and leaving the object arm untouched
\cite{gatti:2003}. 
Along the same line, recent experimental works \cite{howell:2004,bennink:2004,d'angelo:2004},
pointed out a momentum-position realization of the EPR paradox using entangled photon pairs produced by PDC. 
Based on this result, the Authors of \cite{bennink:2004,d'angelo:2004}
argued  that in ghost imaging schemes entangled photons allows to achieve a better spatial resolution than any classically correlated beanms,
and in particular set a lower limit to the product of the resolutions of
the images formed in the near and in the far-field of a given, single, classical source.\\
This claim is quite in contrast with the theoretical analyses of\cite{gatti:2004}, where some of us  proposed a particular source of classically correlated beams, that can emulate the behaviour of entangled beams in all the relevant aspects of ghost imaging, including the resolution capabilities. 
The proposed scheme relied on the use of two beams obtained  by dividing
on a beam splitter  an intense beam with a thermal-like statistics. 
The two outcoming beams  were shown to have strong,
although completely classical, spatial correlations simultaneously in the near and in the far-field planes.
Thanks to this,  the scheme is able to reproduce all the results of ghost imaging with entangled
beams, and numerical simulations confirmed this. In this paper we
provide the first -- to our knowledge -- experimental evidence of the theoretical 
results obtained there.  We show that both the ghost image and the
ghost diffraction results emerge from the correlation measurements, that we may
pass from one  to the other by solely operating on the setup in
the reference arm, and that the product of the resolutions clearly overcomes the limit set in \cite{bennink:2004,d'angelo:2004}.This definitively demonstrates that entanglement is
not necessary for ghost imaging.
\par
\begin{figure}[ht] 
\centerline{    
    \scalebox{.55}{\includegraphics*{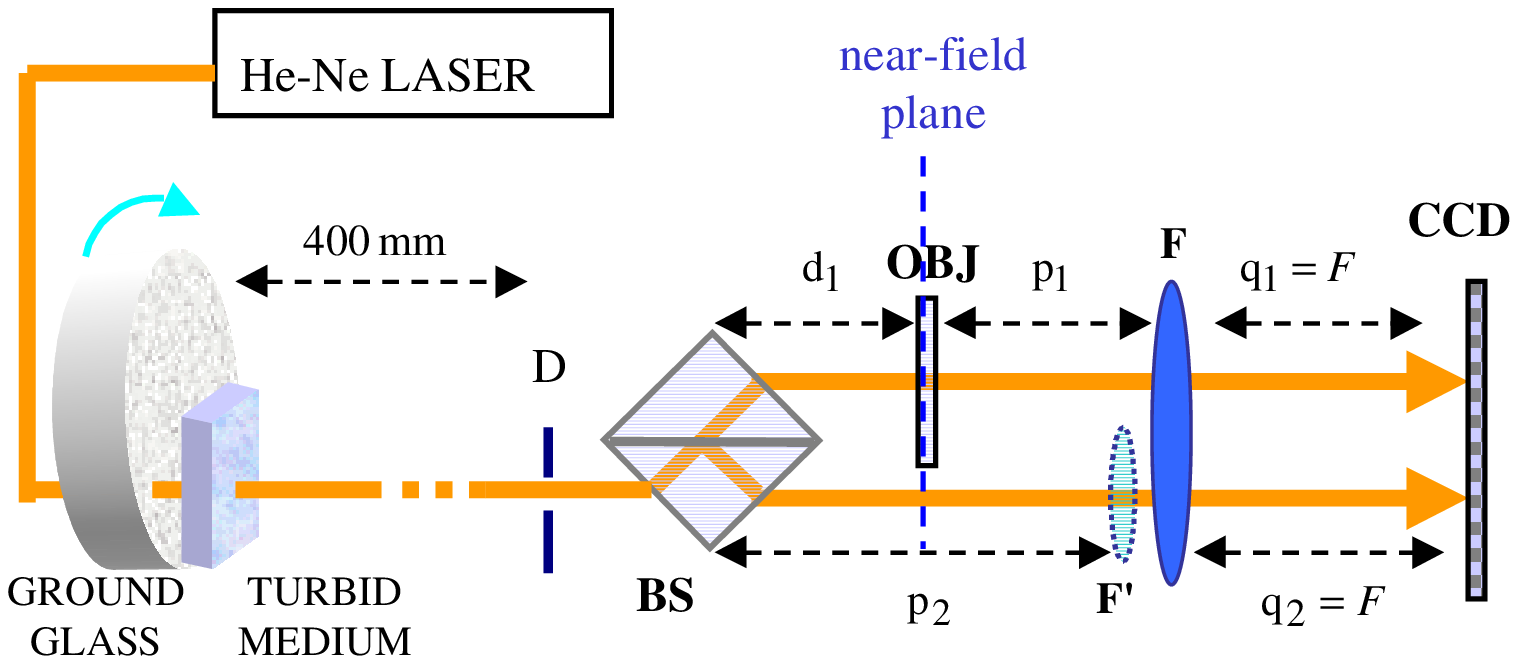}} }
\caption{Scheme of the setup of the experiment (see text for details). }
\label{fig:setup}
\end{figure}

\begin{figure}[ht] 
\centerline{    
    \scalebox{.55}{\includegraphics*{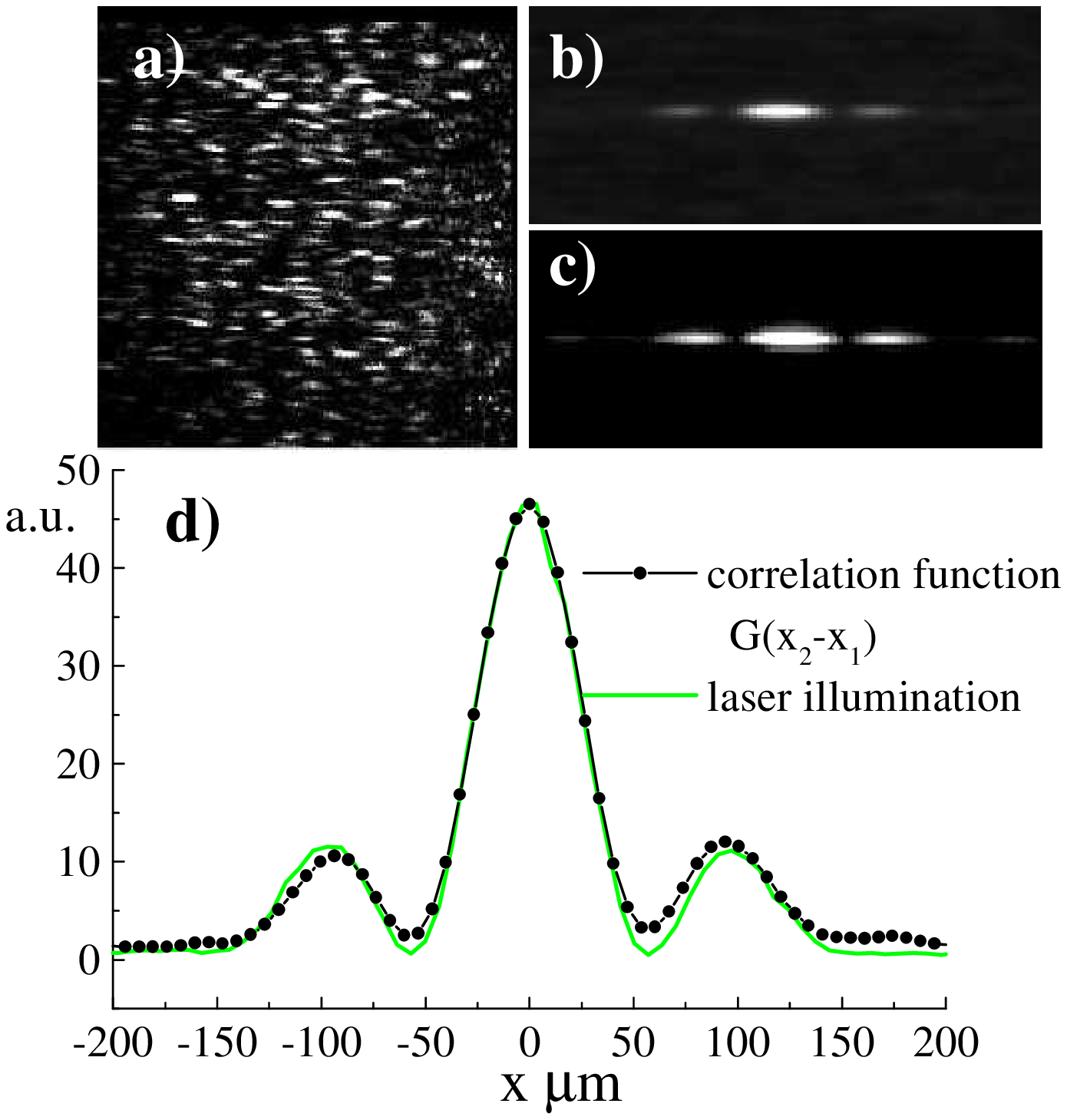}} }
\caption{Recontruction of the diffraction pattern via correlation measurements (Fig.\ref{fig:setup}, with the lens $F^{\prime}$ removed). (a) Single-shot intensity distribution in the object arm, showing no diffraction pattern. (b) Cross-correlation of the object-reference intensity distributions as a function of $\x_2 -\x_1$ (statistics over 500 frames); (c) diffraction pattern observed by shining the laser light on the object. (d) Horizontal sections of the diffraction patterns shown in (b)(dotted)plotted as a function of $x=x_2-x_1$, and (c) (full line)}
\label{fig:diffraction}
\end{figure}

 
The experimental setup is sketched in Fig.\ref{fig:setup}.
 The source of thermal light is provided by a slowly rotating ground 
glass placed in front of a scattering cell containing a higly turbid solution of 
$3\,\mu$m latex spheres. When this is shined with a large collimated 
laser beam ($\lambda = 0.6328\, \mu$m, diameter $D_0 \approx 10$ mm), the 
stochastic interference of the waves emerging from the source produces a 
time-dependent speckle pattern characterized by a correlation time $\tau_{coh}$ on the order of  1s. The latter is associated to particle motion and can 
be tuned by varying  the turbidity of the solution. A small portion of the speckle pattern is selected by a $D=3$ mm 
diaphragm at a distance $z_0=400$ mm from the thermal source. The diaphragm 
realizes an angular selection of the pattern allowing the formation of an 
almost collimated speckle beam  characterized by chaotic statistics\cite{goodman} and
by speckles of size $\Delta x \approx \lambda z_0/D_0 \approx 25 \,\mu$m \cite{goodman}. The speckle beam is separated by the beam splitter (BS) into 
"twin" speckle beams, that exhibit a high  (although classical)  level of spatial correlation. 
The two beams emerging from the BS have slightly non-collinear propagation directions, and  illuminate  two different non overlapping portions of the charged-coupled-device (CCD) camera. The data are 
acquired with an exposure time ($1$ ms) much shorter than $\tau_{coh}$, allowing the recording of high contrast speckle patterns. The frames are 
grabbed at a rate of 1 Hz, so that each data acquisition corresponds to 
uncorrelated speckle patterns. \\
The optical setup of the object arm 1 is fixed. An object, consisting of a thin needle of 160 $\mu$m diameter inside a rectangular aperture 690 $\mu$m wide, is placed in this arm at a distance $d_1$ from the BS. A single lens of focal length $F=80\,$mm is placed after the  object, at a distance $p_1$ from the object and $q_1=F$ from the CCD. In this way the CCD images the far-field plane with respect to the object.
However, the light being incoherent, the diffraction pattern of the object is not visible   (nor its  image, in this setup),  in the intensity distribution  of the object beam on the CCD, as shown by Fig.\ref{fig:diffraction}a.
\begin{figure}[ht] 
\centerline{    
    \scalebox{.55}{\includegraphics*{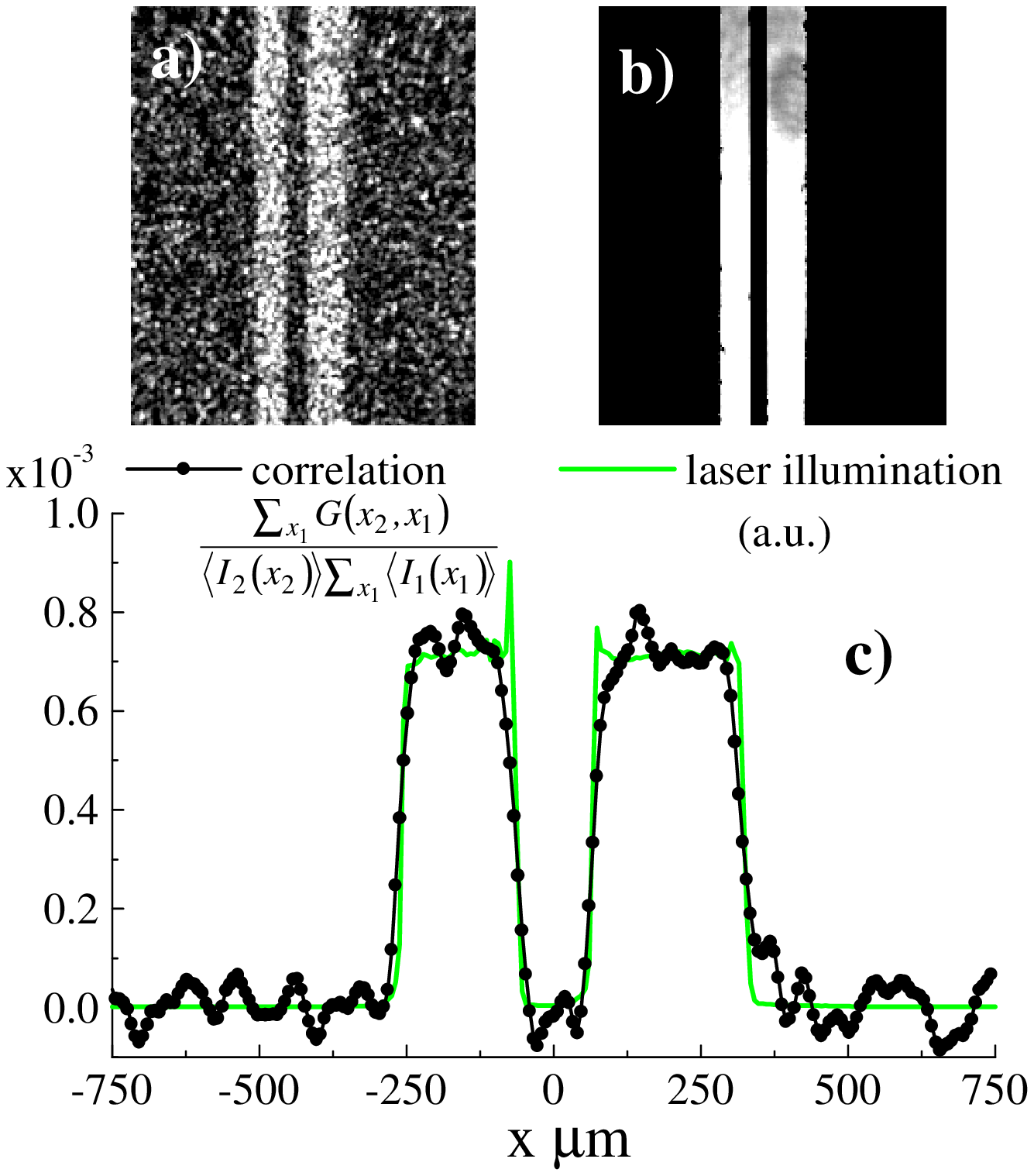}} }
\caption{Recontruction of the object image via correlation measurements  (Fig.\ref{fig:setup}, with the lens $F^{\prime}$ inserted). (a) The intensity distribution of the reference arm is cross-correlated with the total photocounts in the object arm (statistics over 5000 CCD frames); (b) image observed by shining the laser light on the object. (d) The averages of 500 horizontal sections of the images shown in (a)(circles), and (b) (full line) }
\label{fig:image}
\end{figure}  
We consider two different setups for the reference arm 2. 
In the first one, an additional lens of focal length $F^{\prime}$ is  inserted
in the reference arm  immediately before the lens $F$. The total focal length $F_2$ of the two-lens system is smaller
than its distance from the CCD $q_2=F$, being $ \frac{1}{F_2}\approx\frac{1}{F}+ \frac{1}{F^{\prime}}$. 
This allowed us to locate the position of the plane conjugate to
the detection plane, by temporarily inserting the object in the reference arm and illuminating it with the laser light. After having determined the position of the object that produced a well focussed image on the CCD (image shown in Fig.\ref{fig:image}b), the object was translated in the object arm. 
The distances in the reference arm approximately obey a thin lens equation of the form 
${1}/{(p_2-d_1)} + {1}/{q_2} \approx {1}/{F_2}$ 
\footnote{This is only approximately true because the two-lens system is equivalent to a thick lens rather than a thin lens.}.
The data of the intensity distribution of the reference arm are acquired and each pixel is correlated with the total photon counts of the object arm (i.e. with the sum of the photocounts over  a proper set  of the pixels in the object arm), which corresponds to having a  "bucket" detector in arm 1.
Averages performed over few thousands of data acquisitions are enough to show a well-resolved reproduction of  the image of the needle. This is shown in Fig.\ref{fig:image}a and can be compared with the image obtained by illuminating the object with the laser light (Fig.\ref{fig:image}b). Fig.\ref{fig:image}c plots
the horizontal sections of the needle
image as obtained via correlation (circles) and with the laser illumination (light full line), after averaging over 500 pixels in the vertical direction(exploiting the vertical symmetry of the image). The spatial resolution shown by correlated imaging with incoherent light is comparable with that obtained via coherent illumination.  \\
In the second setup, without changing anything in the object arm, 
the lens $F^{\prime}$ is removed from the scheme of Fig.\ref{fig:setup}, so that the reference beam passes through  a single lens of focal $F$, placed at a distance $q_2 =F$ from the CCD.
The spatial cross-correlation  function of  the reference and test arm 
intensity distributions
 is calculated by making averages over few hundreds of  independent data acquisitions, and shows a sharp reproduction of the diffraction pattern of the object (Fig.\ref{fig:diffraction}b). This is comparable with the diffraction pattern obtained  by illuminating the object with the laser light
(Fig.\ref{fig:diffraction}c) (hence, by simply removing the scattering media and observing the light distribution of the object arm on the CCD). Horizontal sections of these diffraction patterns are plotted in Fig. \ref{fig:diffraction}d, and shows basically that the spatial correlation measurement of the speckle beams offers a perfect reproduction of
the diffraction pattern of this object. Notice that at difference from the ghost image experiment a spatial average is also involved, which improves the covergence rate as described in detail in \cite{bache:2004a}.\\
Thus, in this way a high-resolution reconstruction of both the image and the diffraction pattern of the object have been performed by only operating on the optical setup of the reference arm and by using a single classical source.
\par
The basic theory behind the setup shown in Fig.~\ref{fig:setup} has
been explained in detail in Ref.~\cite{gatti:2004}. The
starting point is the input-output relations of a beam splitter 
\begin{equation}
b_1 (\x )= t a(\x) + r v (\x) \, ,
b_2 (\x )= r a (\x)+ t v(\x) 
\label{BS} 
\end{equation}
where $b_1$ and $b_2$ are the object and reference beams emerging from the BS, $t$ and $r$ are the transmission and reflection coefficients of
the BS, $a$ is the speckle field and $v$ is a vacuum field
uncorrelated from $a$. The state of $a$ is a thermal mixture,
characterized by a Gaussian field statistics, in which any correlation
function of arbitrary order is expressed via the second order
correlation function
\begin{equation}
\Gamma(\x,\xp)=
\langle a^\dagger (\x) a (\xp) \rangle
\label{gamma}
\end{equation}
The object
information is extracted by measuring the spatial correlation function
of the intensities $\langle I_1 (\x_1) I_2 (\x_2) \rangle$,
where 
$ I_i (\x_i)$ are operators associated to the number of photocounts 
over the CCD pixel located at  $\x_i$ in the i-th beam,  
and from this calculating the correlation function of intensity
fluctuations
\begin{equation}
G(\x_1, \x_2) = \langle I_1 (\x_1) I_2 (\x_2) \rangle - \langle I_1 (\x_1)\rangle \langle I_2 (\x_2) \rangle \; .
\label{eq6}
\end{equation}
The main result obtained in \cite{gatti:2004} was
\begin{eqnarray}
G(\x_1, \x_2) = |tr|^2
\left| \int {\rm d} \xp_1
\int {\rm d} \xp_2  h_1^* (\x_1, \xp_1) h_2 (\x_2, \xp_2) \Gamma(\x,\xp)
\right|^2 \; ,
\label{eq12}
\end{eqnarray}
where $h_1 $ and $h_2$ are the impulse response function describing the optical setups in the two arms.
Let us consider the scheme shown in Fig.\ref{fig:setup}. The beams $b_1$ and 
$b_2$ follow identical paths from the BS to the object plane, which we shall refer to as the {\em near-field plane}. Due to the linearity of the BS transformation, the two beams at the near-field plane can be expressed via the same transformation as (\ref{BS}), with the $a$ replaced by the  speckle field which has propagated from
the source to the near-field. In other words, our setup is equivalent to a setup where the physical BS is replaced by a point-like 
beam splitter performing the tranformation (\ref{BS}) immediately before the object. \\
The setup of
arm 1 is fixed, and by denoting with 
$T(\x)$ the object transmission function, 
$h_1(\x_1,\xp_1)=(\im\lambda F)^{-1} e^{-\frac{2\pi\im}{\lambda F}
  \x_1 \cdot \xp_1}T(\xp_1)$.
In the ghost image setup (Fig.\ref{fig:setup} with the lens $F'$ inserted), the detection plane of the reference arm
is the conjugate plane with respect to the near-field plane. Hence, apart from inessential phase factors
$h_2 (\x_2,\xp_2) = m\delta
(m \x_2+\xp_2) $, where $m \approx 1.2$ is the magnification factor of the two-lens system.
Inserting this in Eq. (\ref{eq12}):
\begin{eqnarray}
G (\x_1, \x_2)\!\! &\propto&\!\!
\left|
\int {\rm d} \x_1'
 \Gamma_n (\x_1', -m \x_2) 
T^*( \x_1') e^{\im \frac{2\pi}{\lambda f}\x_1 \cdot \x_1'}
\right|^2
\label{image1} \\
\!\!&\approx& \!\!
\left|  T \left( -m \x_2  \right)
\right|^2 
\left|
\int {\rm d} \x_1'
 \Gamma_n (\x_1', -m \x_2) 
e^{\im \frac{2\pi}{\lambda f}\x_1 \cdot \x_1'}
\right|^2\; ,
\label{image2}
\end{eqnarray}
where $\Gamma_n (\x,\xp)$ is the second order field correlation function (\ref{gamma}) of the speckle beam in the near-field.
The second line of Eq.(\ref{image2}) was derived under the assumption that the
smallest scale over which the object changes is larger   than the length over which  $\Gamma_n (\x-\xp)$ decays, which we shall refer to as the {\em near-field coherence length} $\Delta x_n$. In general the result of a measurement of $G$ in this setup is a convolution of the object transmission function with the near field correlation function $\Gamma_n$, so that that $\Delta x_n$  sets the spatial resolution for the reconstruction of the image.
Note that in the ghost image experiment a  bucket detection scheme is employed in arm 1, so what we observe in Fig.\ref{fig:image}a is $\int {\rm d} \x_1 G(\x_1,\x_2)$, which makes the imaging incoherent \cite{bache:2004a}.
\par
In the ghost
diffraction experiment we simply remove the lens $F'$  from the setup of Fig.\ref{fig:setup}, so that the detection plane of beam 2 is in the focal plane of the lens $F$. Hence, $h_2(\x_2,\xp_2)= (\im\lambda F)^{-1} e^{- \frac{2\pi\im}{\lambda
    F} \x_2 \cdot \xp_2 }$. From Eq.(\ref{eq12}), we obtain
\begin{eqnarray}
G (\x_1, \x_2)\!\! &\propto&\!\!
\left|
\int {\rm d} \vec{\xi}
 \Gamma_f (\vec{\xi}, \x_2) 
\tilde{T}^* \left[ (\x_1-\vec{\xi}) \frac{2\pi}{\lambda F} \right] 
\right|^2
\label{diff1} \\
\!\!&\approx& \!\!
\left|  \tilde{T} \left[ (\x_1-\x_2) \frac{2\pi}{\lambda F} \right] 
\right|^2 
\left|
\int {\rm d} \vec{\xi}
 \Gamma_f( \vec{\xi},\x_2) 
\right|^2\; ,
\label{diff22}
\end{eqnarray}
where $\tilde T (\q) = \int \frac{{\rm d} \x}{2\pi} e^{-\im \q \cdot
  \x} T(\x)$ is the amplitude of the diffraction pattern from the
object. $\Gamma_f(\x,\xp)$ is the second order correlation function (\ref{gamma}) of the speckle field in the far-field plane, as observed in the focal plane of the lens $F$; its correlation length $\Delta x_f$ (the {\em far-field coherence length})  sets the limit for the spatial resolution of the diffraction pattern reconstruction.\par
\begin{figure}[hb] 
\centerline{    
    \scalebox{.50}{\includegraphics*{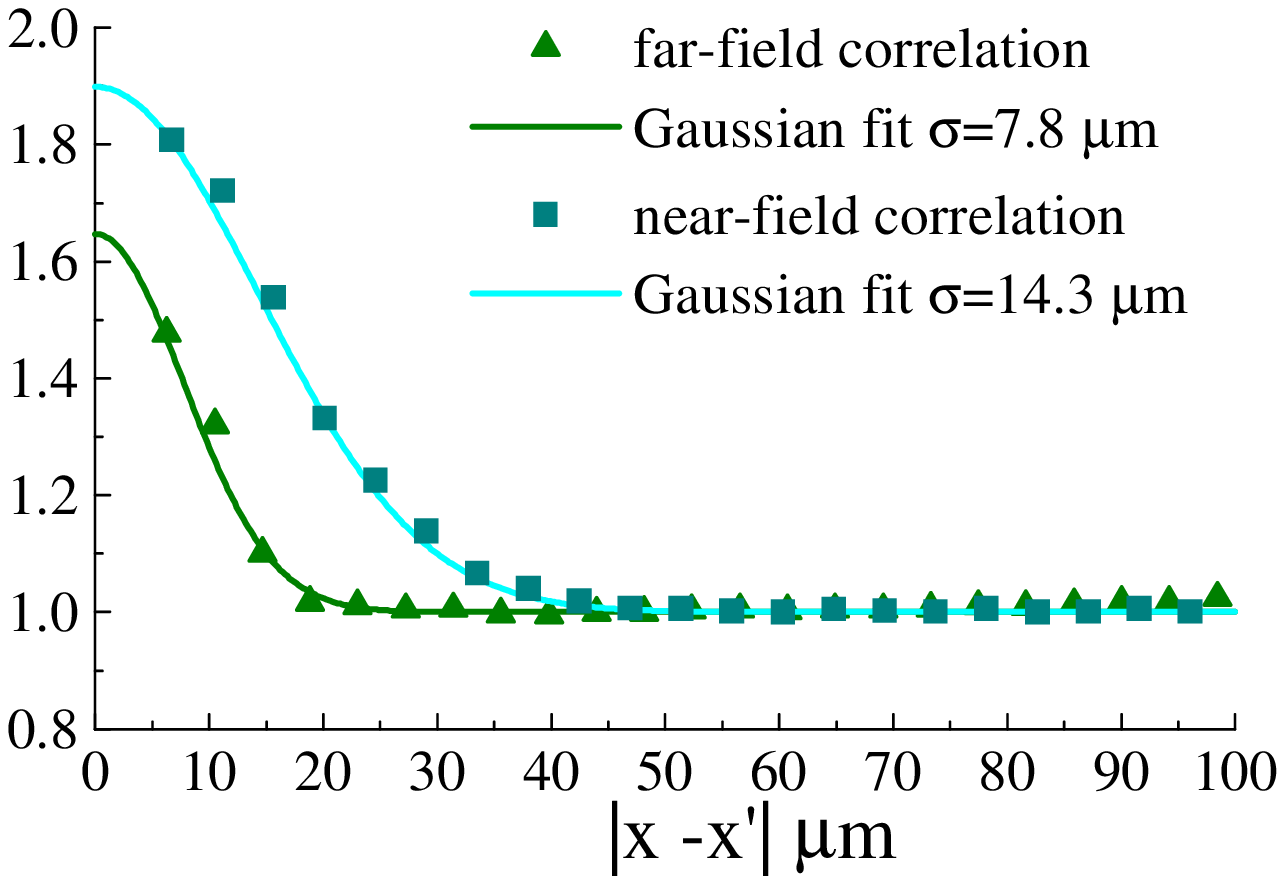}} }
\caption{Fourth order auto-correlation function in the near field plane(squares) and in the far-field plane(triangles). The full lines are Gaussian fits of the correlation peaks, and the data have been normalized to the baseline values.}
\label{fig:corr}
\end{figure}

Relevant to the resolution of the ghost imaging and ghost diffraction schemes are hence the spatial coherence properties 
of the speckle field in the near field immediately before the object,
and in the far-field plane. 
These can be investigated by measuring the fourth-order correlation functions, in the absence of the object. In the near-field plane, we measured the spatial auto-correlation
function of the reference beam  $\langle I_2(\x) I_2 (\xp)\rangle$
in the setup with the lens $F'$ inserted, so that the 
reference beam recorded by the CCD is the (demagnified) image of the near-field. This is plotted  in Fig.\ref{fig:corr} (squares) as a function of $|\x -\xp|$.
Neglecting the small shot noise contribution at $\x=\xp$, by using Siegert formula for Gaussian field statistics and the BS tranformation (\ref{BS}),  
\begin{eqnarray}
\langle  I_2(\x) I_2 (\xp)\rangle&-& 
\langle  I_2(\x)\rangle \langle I_2 (\xp)\rangle \nonumber \\
&=&
\left| \langle b_2^\dagger (\x) b_2 (\xp) \rangle \right|^2
= |r|^4  \left| \Gamma_n (m \x ,m \xp)
\right|^2 \; ,
\label{gamman}
\end{eqnarray}
The baseline in Fig.\ref{fig:corr} corresponds to the product of mean intensities, while 
the narrow peak at  $\x=\xp$ is the r.h.s of (\ref{gamman}), and reflects the spatial coherence properties of the field. By making a Gaussian fit of this peak we obtained a variance $\sigma_n=14.3 \, \mu$m. From this we can infer the coherence length in  the near-field plane, as $\Delta x_n \approx 2 m \sigma_n =34.3\,\mu$m. 
The triangles in Fig.\ref{fig:corr} plot the intensity correlation function
in the far-field plane, obtained by measuring the auto-correlation function of beam 1 in the focal plane of the lens $F$. The narrow peak in this plot is now $|t|^4 \left| \Gamma_f(\x,\xp) \right|^2$ and reflects the spatial coherence properties of the speckles in the far-field plane. Its variance calculated by means of a Gaussian fit
is $\sigma_f=7.8 \, \mu$m, from which we infer a far-field coherence length $\Delta x_n= 2 \sigma_f=15.6 \,\mu$m. This in turn corresponds to a spread in transverse wave vectors $\Delta q= \frac{2 \pi}{\lambda F} \Delta x_f= 1.93 \times 10^{-3}\, \mu m^{-1}$\\
In the spirit of the EPR argument brought forward by \cite{bennink:2004,d'angelo:2004}, we find a product of the near-field and far-field resolutions 
\begin{equation}
\Delta x_n \Delta q = 0.066
\end{equation}
which is by far smaller than the limit of unity imposed by EPR-like arguments in\cite{bennink:2004,d'angelo:2004}, and is at least four times smaller than the results there reported, where entangled photons were used to retrieve the ghost image and the ghost diffraction pattern.\\
Notice that there is  no violation of any EPR bound for such a separable  mixture. In fact, in any plane, the conditional probability of detecting a photon  at position $\x_2$ in beam 2 given that a photon was detected at position $\x_1$ in beam 1 is:
\begin{eqnarray}
P(\x_2|\x_1) \propto \frac {\langle  I_2(\x_2) I_1 (\x_1)\rangle} {\langle I_1 (\x_1)\rangle} =\langle  I_2(\x_2) \rangle + |tr|^2\frac{\left| \Gamma (\x_1,\x_2\right|^2 } {\langle I_1 (\x_1)\rangle}  \; .
\label{conditional} 
\end{eqnarray}
Hence, given that photon 1 was detected at $\x_1$ there is a large probability of detecting a photon anywhere in beam 2 (the first term at r.h.s. of Eq. (\ref{conditional})), 
with a narrow probability peak localized at $\x_2 = \x_1$. Thanks to the first term,  there is no non-locality ingredient in the correlation of the classical speckle beams, as obviously must be. However, the crucial point is that the resolution of imaging schemes based on correlation measurements depends only on the widths 
of the  peaks in Fig. \ref{fig:corr}, and that their product in the near and in the far field are not bounded by any EPR-like inequality. As a matter of fact the near and the far field coherence lengths, corresponding roughly to the size of the speckles in the near and in the far-field, are independent quantities. In the near field, the size of the speckle depends on the laser diameter $D_0$, and from the total optical path length $z$ from the source to the near-field plane,  $\Delta x_n \propto \lambda z/D_0$ \cite{goodman}. The pinhole $D$ being rather close to the near-field plane, we checked that the speckle size in this plane did not depend on the pinhole aperture. The pinhole aperture instead influences the size of the speckles in the far field, this being  roughly given by $\Delta x_f \propto \lambda F/D$\cite{goodman}. 
Using the values of our setup, we find $\Delta x_n \approx 30 \, \mu m$ and  
$\Delta x_f \approx 17 \, \mu m$, in good agreement with the values estimated 
from the correlation (Fig. \ref{fig:corr}).
\par
In conclusion, we have reported on an experimental demostration of high resolution  ghost imaging and ghost diffraction by using a single source of  classically correlated thermal light. The distributed imaging character of our experiment is evident from the fact that the information on the object spatial distribution can be processed  by only acting on the reference beam 2(e.g. passing from the image to the diffraction pattern). The product of resolutions 
of the ghost image and the ghost diffraction schemes is well below the limit that was claimed by former literature to be achievable only with entangled beams. This could be further improved by optimizing the scheme (e.g. with a larger pinhole, a larger laser diameter) which was constrained in our experiment by the need of using a single CCD and a low-power laser.
This definitely proves  the claim set forth in \cite{gatti:2004}, that the  only advantage of entanglement with respect to classical correlation lies in
the better visibility of information. This  might be useful in very high sensitivity measurement or in quantum information schemes, where e.g. the information  need to be hidden to a third party, but does not give any evident practical advantage when attempting to process information from a macroscopic classical object, as in the experiment shown here.
\par

\acknowledgments{This work was carried out in the framework of the 
  project QUANTIM of the EU, of the PRIN project of MIUR "Theoretical study of novel devices based on quantum entanglement", and of the INTAS project "Non-classical light in quantum imaging and continuous variable quantum channels". M.B. acknowledges financial support from the Danish Technical
  Research Council (STVF) and the Carlsberg Foundation.}
\bibliographystyle{C:/texmf/tex/latex/revtex4/apsrev}
\bibliography{d:/Projects/Bibtex/literature}

\end{document}